\documentclass[prl,superscriptaddress,showpacs,twocolumn]{revtex4}
\usepackage{graphicx}
\usepackage{dcolumn}
\usepackage{bm}
\usepackage{SIunits}

\renewcommand{\vec}{\bm}
\newcommand{\miller}[3]{\ensuremath{(#1, #2, #3)}}
\newcommand{\direction}[3]{\ensuremath{[#1, #2, #3]}}
\renewcommand{\deg}{\ensuremath{^{\circ}}}

\begin{document}

\title{Nature of the magnetic order and origin of induced ferroelectricity in TbMnO$_3$}

\author{S. B. Wilkins}
\affiliation{Department of Condensed Matter Physics and Materials Science, Brookhaven National Lab, Upton, New York 11973-5000, USA}
\author{T. R. Forrest}
\affiliation{London Centre for Nanotechnology, University College London, Gower Street,
London WC1E 6BT, UK}
\author{T.A.W. Beale}
\author{S.R. Bland}
\affiliation{Department of Physics, Durham University, South Road, Durham, DH1 3LE, UK}
\author{H.C. Walker\footnote{Present address: European Synchrotron Radiation Facility, Bo\^\i te Postal 220, F-38043 Grenoble CEDEX, France}}
\affiliation{London Centre for Nanotechnology, University College London, Gower Street,
London WC1E 6BT, UK}
\author{D. Mannix}
\affiliation{Institut N\'eel, CNRS-UJF, BP166, 38042 Grenoble, France}
\author{F. Yakhou}
\affiliation{European Synchrotron Radiation Facility, Bo\^\i te Postal 220, F-38043 Grenoble CEDEX, France}
\author{D. Prabhakaran}
\author{A.T. Boothroyd}
\affiliation{Department of Physics, University of Oxford, Clarendon Laboratory, Parks Road,
Oxford OX1 3PU, UK}
\author{J.P. Hill}
\affiliation{Department of Condensed Matter Physics and Materials Science, Brookhaven National Lab, Upton, New York 11973-5000, USA}
\author{P.D. Hatton}
\affiliation{Department of Physics, Durham University, South Road, Durham, DH1 3LE, UK}
\author{D.F. McMorrow}
\affiliation{London Centre for Nanotechnology, University College London, Gower Street,
London WC1E 6BT, UK}

\date{\today}
\begin{abstract}
The magnetic structures which endow TbMnO$_3$ with its multiferroic properties have been reassessed on the basis of a comprehensive soft x-ray resonant scattering (XRS) study. The selectivity of XRS facilitated separation of the various contributions  (Mn $L_2$ edge, Mn $3d$ moments; Tb M$_4$ edge, Tb $4f$ moments), while its variation with azimuth provided information on the moment direction of distinct Fourier components.  When the data are combined with a detailed group theory analysis, a new picture emerges of the ferroelectric transition at 28~K. Instead of being driven by the transition from a collinear to a non-collinear magnetic structure, as has previously been supposed, it is shown to occur between two non-collinear structures.
\end{abstract}
\maketitle

Recently, considerable interest has been focused on a class of materials, {\it multiferroics}\cite{Khomskii:2009p1669}, that display strongly coupled magnetic and ferroelectric order parameters, as exemplified by the ground breaking study on TbMnO$_3$\cite{Kimura:2003p1023}. Of particular significance is the fact that ferroelectricity results from phase transitions between different magnetic structures\cite{Kimura:2003p1023, Hur:2004p6, Kenzelmann:2005p19,Lawes:2005p1909}. This fact has facilitated a number of studies in which the ferroelectric state can be controlled by the application of an external magnetic field\cite{Hur:2004p6,aliouane:020102,Strempfer:2008p1607}. Developing a microscopic understanding of these effects represents a considerable challenge from a fundamental physics perspective, while their exploitation may lead to novel devices\cite{Khomskii:2009p1669}.

In the case of TbMnO$_3$, ferroelectricity occurs below 28~\kelvin,  concomitant with a phase transition between two magnetic phases\cite{kimura:224425}. Using neutron diffraction, Kenzelmann {\it et al.}\cite{Kenzelmann:2005p19} have proposed that at this temperature the Mn magnetic moments undergo a collinear to non-collinear transition, to form a cycloid which removes an inversion centre resulting in a ferroelectric polarization. Various theoretical approaches\cite{Katsura:2005p1671, mostovoy:067601} inspired by this insight predict that the ferroelectric polarization $\vec{P}$ is given by:
\begin{equation}
\vec{P} \propto \vec{r}_{ij} \times \vec{S}_i \times \vec{S}_j
\label{eq:P}
\end{equation}
where $\vec{r}_{ij}$ is the vector connecting two nearest-neighbor spins $\vec{S}_i$ and $\vec{S}_j$.
Clearly, of central importance in our understanding of this class of multiferroic is obtaining a complete and accurate description of the magnetic structures they possess.

We measure reflections in addition to the $(0,\tau,1)$ type observed in Ref. \cite{Kenzelmann:2005p19}, by soft x-ray resonant scattering. This allows us to definitively measure the Fourier components with element specific sensitivity, not possible in previous neutron diffracton studies. We find that the transition associated with ferroelectricity is not one from a collinear to non-collinear cycloidal magnetic structure, as has been previously reported \cite{Kenzelmann:2005p19}, but that the transition actually corresponds to one between two non-collinear magnetic structures, where below 28~\kelvin\ there exists a cycloidal component. This result is found by considering all observed reflections to be Fourier components of a single magnetic structure. 


\begin{figure}[t!]
\centering
\includegraphics[width=0.75\columnwidth]{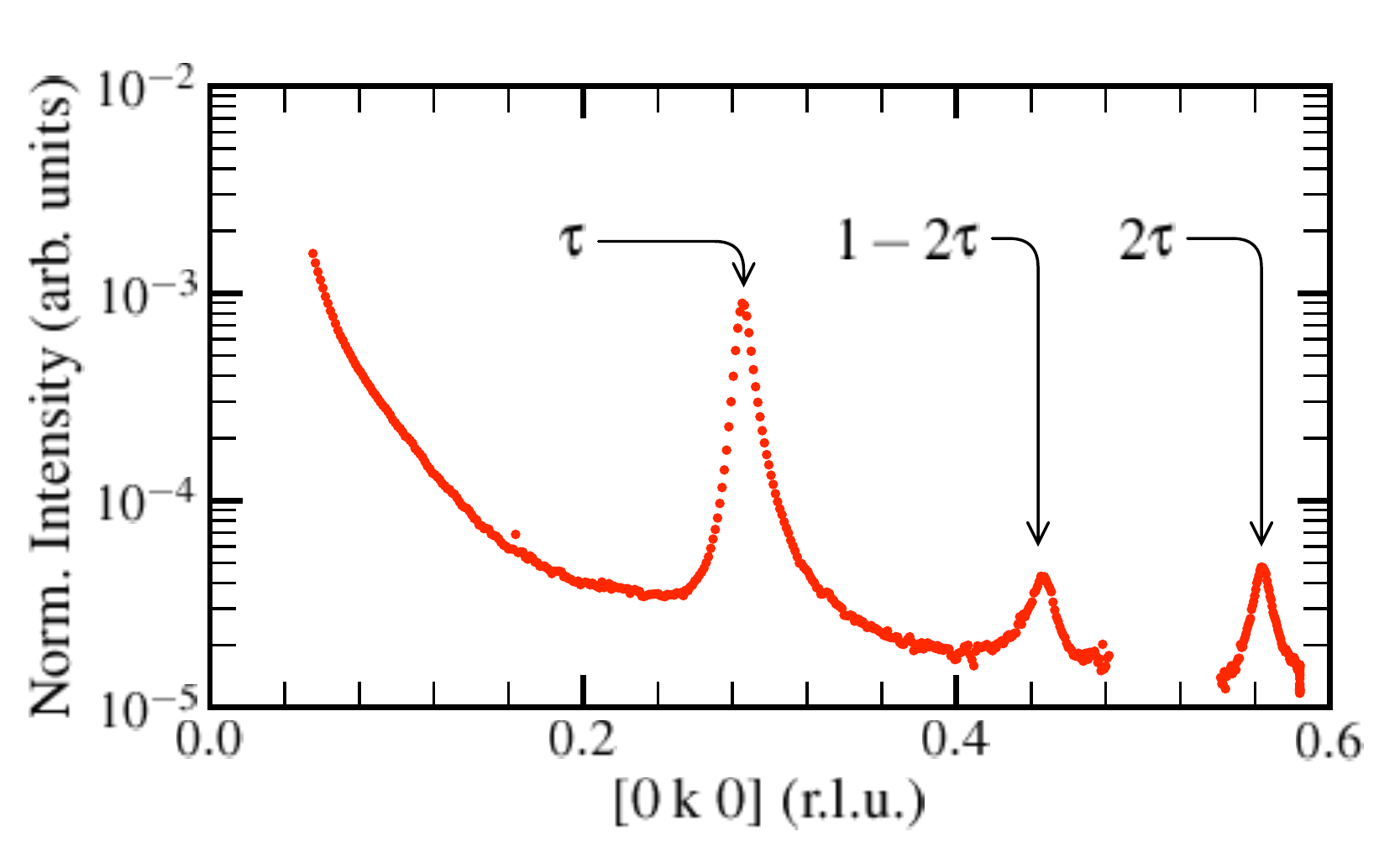}
\caption{(color online) Scan along the $[010]$ direction in reciprocal space at an incident photon energy corresponding to the L$_2$ edge of Mn.}
\label{fig:longq}
\end{figure}


Neutron diffraction measurements\cite{Kajimoto:2004p1616,Kenzelmann:2005p19, Quezel:1977p1673} have established that there are three magnetic transitions in TbMnO$_3$. Below $T_\mathrm{N1} = 42.5$~\kelvin\ the Mn$^{3+}$ ions develop long range order. In the work of Kajimoto {\it et al}.,\cite{Kajimoto:2004p1616} magnetic superlattice reflections were observed at \miller{0}{\tau}{l} and \miller{0}{1-\tau}{l} type positions with $l$ integer and $\tau = 0.27$. At $T_\mathrm{N2} = 28$~\kelvin\ there is a second magnetic transition which coincides with the appearance of a spontaneous electric polarization along the $c$-axis. Finally, the Tb ions order below a third transition at $T_\mathrm{Tb} = 7$~\kelvin. Kenzelmann {\it et al}.,\cite{Kenzelmann:2005p19} developed an appealing model of the magnetic structure whereby above $T_\mathrm{N2}$ the magnetic moments on the Mn ions are polarized along the $b$-axis in an incommensurate, collinear fashion, and argued that at $T_\mathrm{N2}$ the magnetic moments on the Mn sites rotate to form an elliptical cycloid in the $bc$-plane. This non-collinear structure causes a finite ferroelectric polarization along the $c$-axis following Eq.~\ref{eq:P}. Hard x-ray scattering experiments have been performed on TbMnO$_3$\cite{Mannix:2007p1566, prokhnenko:177206, Voigt:2007p1608}, and report both non-resonant and resonant magnetic satellites at \miller{0}{\tau}{l}, \miller{0}{1-\tau}{l} type locations. However, for these experiments the resonances used do not provide direct information on the localised Mn $3d$ or Tb $4f$ states which are of principal interest for understanding the magnetic order \cite{Usuda:2004p1677, Kuzushita:2006p218,Mannix:2001p1678}. The soft XRS experiments reported here overcome this limitation by probing these states directly \cite{Wilkins:2003p1841,Beale:2007p1707,Wilkins:2003p231,Thomas:2004p1680, Dhesi:2004p1681}.

\begin{figure}
\centering
\includegraphics[width=0.75\columnwidth]{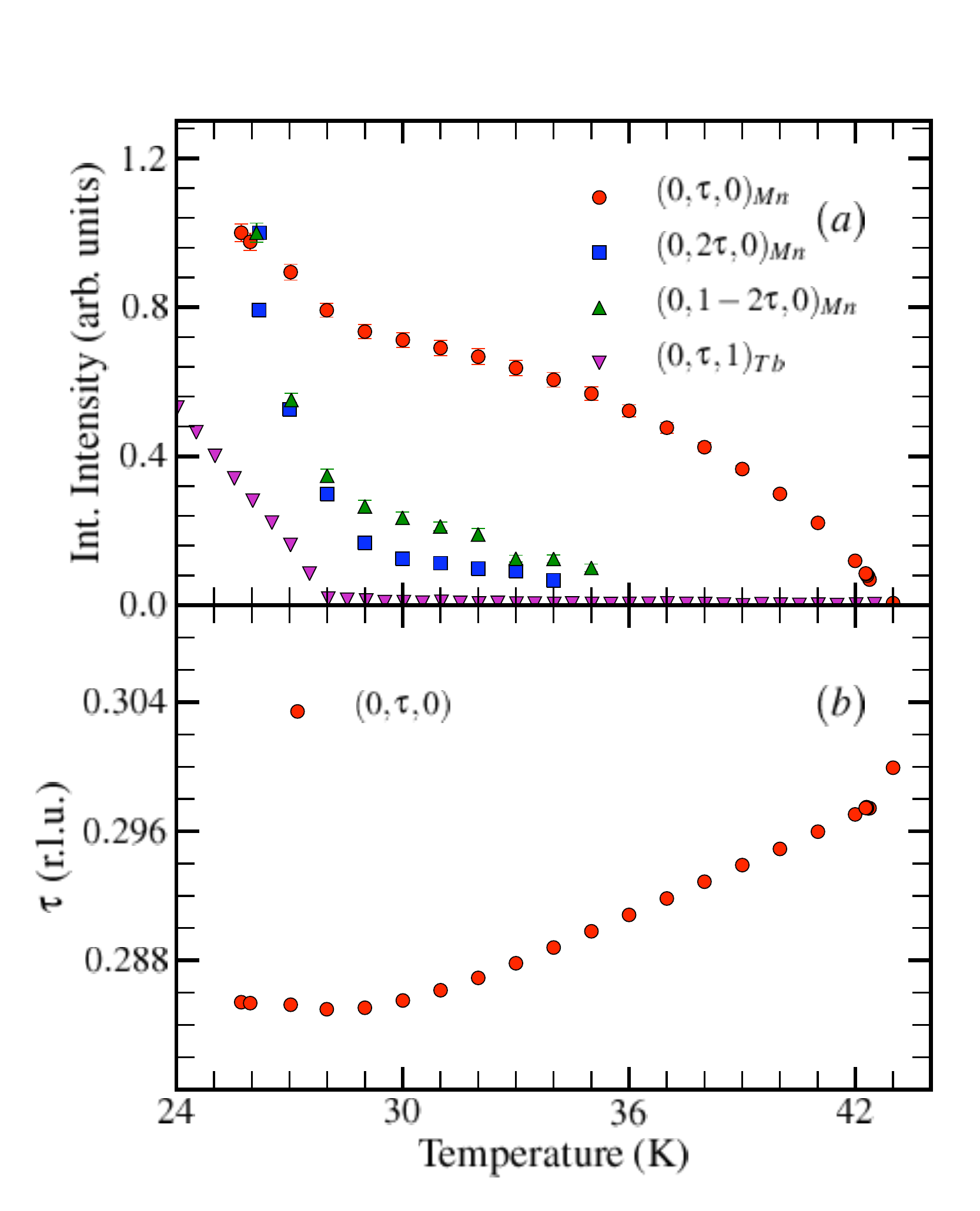}
\caption{(color online) (a) Temperature dependence of the integrated intensity of the $(0, \tau, 0)$, $(0, 2\tau, 0)$, and $(0, 1-2\tau, 0)$ measured at the Mn $L_2$ edge and the $(0, \tau, 1)$ measured at the Tb M$_4$ edge. (b) Temperature dependence of the incommensurability $\tau$ measured from the peak position of the $(0, \tau, 0)$ measured at the Mn $L_2$ edge. }
\label{fig:tdep}
\end{figure}


Experiments were carried out on beamlines 5U1 at the SRS, Daresbury Laboratory, UK and ID08, European Synchrotron Radiation Facility, France. Two crystals of TbMnO$_3$ were used, cut from the same boule as crystals used in previous experiments\cite{Mannix:2007p1566,Forrest:2008p42205}. Faces were cut such that on one crystal the surface normal was \direction{0}{1}{0} and on the other so that the \miller{0}{0.28}{1} reflection was surface normal. The faces were polished to a 0.05~\micro\meter\ finish. Unless otherwise noted, the incident photon energy was tuned to the $L_2$ edge of Mn, to select the magnetic scattering from the manganese. The polarization of the incident x-rays was chosen to be perpendicular to the scattering plane ($\sigma$ polarization). 

Figure~\ref{fig:longq} shows a scan taken at 20~K along the \direction{0}{1}{0} direction in  reciprocal space from the crystal with \direction{0}{1}{0} surface normal. Three superlattice reflections are visible at, \miller{0}{\tau}{0}, \miller{0}{2\tau}{0} and \miller{0}{1-2\tau}{0} with $\tau \approx 0.28$. All three reflections were found to resonate at the $L_2$ and $L_3$ edges of Mn and were not visible off resonance. We assign the \miller{0}{\tau}{0} reflection to incommensurate ordering of the magnetic moments on the Mn sites and the \miller{0}{2\tau}{0} to be the second harmonic from the resonant x-ray cross section\cite{HILL:1996p89, HANNON:1988p46}. The \miller{0}{1-2\tau}{0} reflection corresponds to the second harmonic from the \miller{0}{1-\tau}{0} reflection. (Note: the fundamental  \miller{0}{1-\tau}{0} reflection is not visible at the Mn $L_2$ or $L_3$ edges as this wavevector is outside the Ewald sphere.)

Upon cooling the sample with the \miller{0}{0.28}{1} reflection surface normal, and tuning the photon energy to the Tb $M_4$ edge, a superlattice reflection was observed at \miller{0}{\tau}{1} with $\tau \approx 0.28$ below 28~\kelvin. This reflection is not accessible at the Mn $L$-edges, but was visible at the Tb $M_5$ and $M_4$ edges with no intensity observed off resonance.

\begin{figure}[t!]
\centering
\includegraphics[width=0.75\columnwidth]{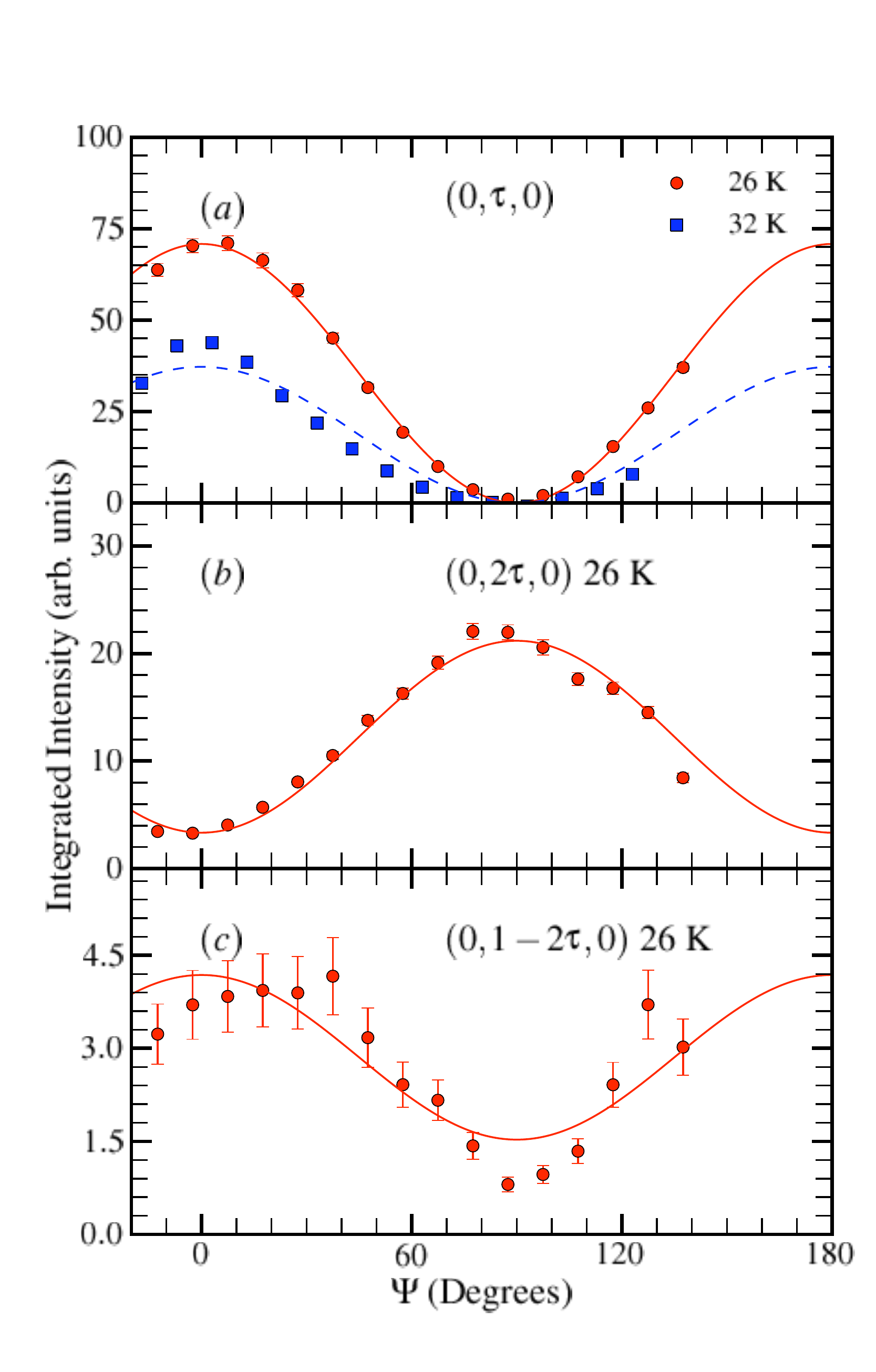}
\caption{(color online) Azimuthal dependence of the (a) $(0,\tau,0)$, (b) $(0,2\tau,0)$ and (c) $(0, 1-2\tau, 0)$ superlattice reflections at the Mn $L_2$ edge in the non-collinear phase at 26~K (red circles). The azimuthal dependence of the $(0,\tau,0)$ at 32~K in the collinear phase is shown for comparison in the top panel (blue squares). The red solid (blue dashed) lines show simulations to the azimuthal dependencies at 26~K (32~K). The origin of the azimuth corresponds to the condition where the $[001]$ direction lies in the scattering plane.}
\label{fig:azimuth}
\end{figure}

The temperature dependence of the integrated intensity of the \miller{0}{\tau}{0}, \miller{0}{2\tau}{0} and \miller{0}{1-2\tau}{0} wavevectors at the Mn $L_2$ edge is shown in Fig.~\ref{fig:tdep}(a). Here a clear transition is seen at $T_\mathrm{N1}$, below which the intensity of the \miller{0}{\tau}{0} grows. At $T_\mathrm{N2}$, a further transition is observed at which the intensity of the \miller{0}{\tau}{0} reflection is found to increase more rapidly. Turning to the \miller{0}{2\tau}{0} and \miller{0}{1-2\tau}{0} wavevectors, these are only observed below a temperature of $\approx 35$~\kelvin, at which point they steadily increase until $T_\mathrm{N2}$. Below $T_\mathrm{N2}$, they too increase more rapidly. Figure~\ref{fig:tdep}(b) shows the temperature dependence of $\tau$. Upon cooling below $T_\mathrm{N1}$, $\tau$ decreases in an almost linear fashion until $T_\mathrm{N2}$ where the value stabilizes at approximately $\tau = 0.287$, although it does not lock-in. Identical behavior was seen in $\tau$ from all reflections. Finally, Fig.~\ref{fig:tdep}(a) also shows the integrated intensity of the \miller{0}{\tau}{1} reflection measured at the Tb $M_4$ edge. This reflection is only visible below $T_\mathrm{N2}$. The presence of the Tb resonant signal is an explicit confirmation of a finite Tb moment in the low $T$ phase, as inferred by Kenzelmann {\it et al.}\cite{Kenzelmann:2005p19} and non-resonant x-ray scattering\cite{fabrizi:237205}.

It is possible to determine the magnetic moment direction by utilizing the fact that the scattered intensity is a function of the projection of the moment along the scattered x-ray direction and the scattering geometry\cite{HILL:1996p89}. In a diffraction experiment, one measures the direction of the {\it Fourier component} of the magnetic structure as determined by the $\vec{Q}$-vector chosen. To determine the direction of the Fourier components present at the \miller{0}{\tau}{0}, \miller{0}{2\tau}{0} and \miller{0}{1-2\tau}{0} wavevectors we measured the azimuthal dependence at each wavevector. This is the intensity dependence obtained on rotating the sample around each $\vec{Q}$ vector. No measurement of the scattered polarization was made.

Figure~\ref{fig:azimuth} shows the azimuthal dependence of the \miller{0}{\tau}{0}, \miller{0}{2\tau}{0} and \miller{0}{1-2\tau}{0} reflections as measured at the Mn $L_2$-edge. In the top panel, the \miller{0}{\tau}{0} reflection is shown at 26~\kelvin\ (red circles) and 32~\kelvin\ (blue squares). While the maximum integrated intensity changes between these two temperatures, the functional form is identical, indicating that the direction of this Fourier component does not change across $T_\mathrm{N2}$. This azimuthal dependence can be fitted to the form:
$I_{\tau}(\psi) = A\cos^2\psi$, 
where $A$ is a proportionality constant, and $\psi$ is the azimuthal angle. $\psi = 0$ corresponds to the condition where the $[001]$ direction lies in the scattering plane. We can immediately exclude a $b$-axis direction for this Fourier component, since this is parallel to the $\vec{Q}$ and thus would be invariant under rotations around $\vec{Q}$. The zero in intensity at $\psi = 90^\circ$, (corresponding to the case where the $c$-axis is perpendicular to the scattering plane), implies that the magnetic moment has no projection along the outgoing x-ray at this azimuth, and this means the $a$ and $b$ axis components are strictly zero at this wavevector. Thus this Fourier component is along the $c$-axis. This result is confirmed by the azimuthal dependence of the \miller{0}{2\tau}{0} wavevector at 26~\kelvin\, shown in  Fig.~\ref{fig:azimuth}. This can be fitted to the form:
$I_{2\tau}(\psi) = A^\prime + B^\prime[\sin^4\psi + \sin^2\psi \cos^2\psi]$.
With the assumption that this reflection is the second harmonic from the resonant x-ray cross section and following Ref.~\onlinecite{HILL:1996p89}, this is also consistent with a Fourier component along the $c$-axis. The finite offset, introduced through the $A^{\prime}$ coefficient is due to scattering from the lattice distortion induced by the magnetic order. This is a pure charge resonant signal and has no azimuthal dependence. No scattering was observed off resonance.

The $1-2\tau$ reflection can be fitted to the form: 
$I_{1-2\tau}(\psi) = A^{\prime\prime} + B^{\prime\prime}[\cos^4\psi + \cos^2\psi \sin^2\psi]$. 
This form is similar to that of the \miller{0}{2\tau}{0} wavevector but with a $\frac{\pi}{2}$ phase shift. Thus the Fourier component at \miller{0}{1-2\tau}{0} and therefore at \miller{0}{1-\tau}{0}, is along the $a$-axis. Voigt {\it et al}.,\cite{Voigt:2007p1608} observed both reflections of the \miller{0}{\tau}{0} and  \miller{0}{1-\tau}{0} type at the Tb L-edges where a similar $\frac{\pi}{2}$ phase shift was observed. However, the data of Voigt {\it et al}. only confirms the presence of a polarization of the Tb $5d$ states at this wavevector. Our data presented here prove that these wavevectors correspond to ordering of the Mn ions.

In summary, we conclude that at 26~\kelvin, the magnetic structure possesses Fourier components at  \miller{0}{\tau}{0} and  \miller{0}{1-\tau}{0} which are along the principal $c$ and $a$ axes respectively, in addition to the to the $b$ axis and $c$ axis components reported previously at \miller{0}{\tau}{1} \cite{Kenzelmann:2005p19}.

To explain the presence of these Fourier components, and the previous neutron results\cite{Kenzelmann:2005p19} one might turn to a model of magnetic domains in which different parts of the sample have different ordering wavevectors. Instead, we propose that these reflections arise from one coherent magnetic structure which explains all the observed wave vectors. This is because 1), the incommensurability is the same and has the same temperature dependence at all reflections, and is consistent with that measured by neutrons\cite{Kajimoto:2004p1616,Kenzelmann:2005p19} and non-resonant x-rays\cite{Mannix:2007p1566, Voigt:2007p1608}. This would seem unlikely if the reflections originated from different parts of the sample with different physical characteristics. 2), the \miller{0}{\tau}{0} wavevector shows magnetic transitions at 41~K and at 28~K; that is, it exhibits transitions entirely consistent with those observed at the \miller{0}{\tau}{1} wavevector \cite{Kenzelmann:2005p19} again suggesting they have a common origin. 3) Such a coherent structure is consistent with the symmetry analysis performed by Kenzelmann {\it et al.,}\cite{Kenzelmann:2005p19}. Finally, we note our sample does indeed show the cycloidal ordering below 28~\kelvin, which we observe at the \miller{0}{\tau}{1} wavevector. We can thus rule out the possibility that this sample somehow differs from others and is not multiferroic.

\begin{figure*}
\centering
\includegraphics[width=1.8\columnwidth]{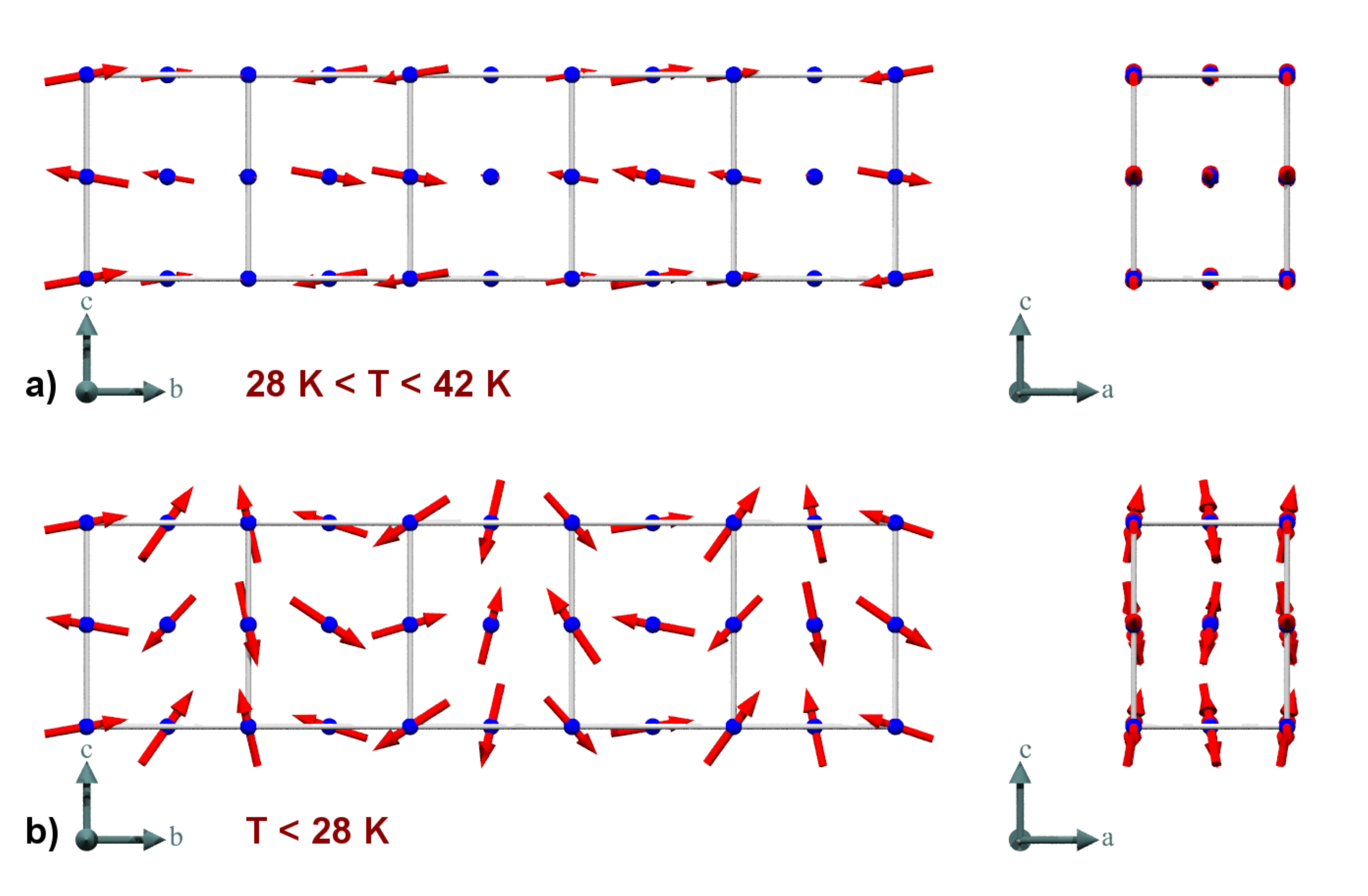}
\caption{(color online) a) Proposed magnetic structure in the region 28~K$ < T < 41$~K projected onto both the crystallographic $b$-$c$ (left) and $a$-$c$ (right) -planes. b) Proposed magnetic structure below 28~K.}
\label{fig:cartoon}
\end{figure*}

We now construct these new magnetic structures. For the high-temperature phase above 28~\kelvin, we start from the model of Kenzelmann {\it et al.,}\cite{Kenzelmann:2005p19}, who found a single $\Gamma_3$ component along the $b$-axis with a wavevector of \miller{0}{\tau}{1}. We add to this the \miller{0}{\tau}{0} $c$-axis Fourier component observed here. This introduces a canting of the spins away from the $b$-axis. Further, this canting must be ferro ordered along the $c$-axis, since this Fourier component is observed at even integer $l$ values. We estimate the canting angle from the neutron intensities of Kajimoto {\it et al.,}\cite{Kajimoto:2004p1616} to be approximately 5\deg\ away from the $b$-axis. This new structure is non-collinear and is shown with this canting angle in Fig.~\ref{fig:cartoon}(a). Note, this structure is consistent with the group theory presented by Kenzelmann {\it et al.,}\cite{Kenzelmann:2005p19}. If we examine the $\Gamma_3$  irreducible representation of the little group $\bm G_k$ of the wavevector \miller{0}{k}{0} from the crystallographic spacegroup {\it Pbnm}, we find that the symmetry operation $m_{xy}$ at  $z = \frac{1}{4}$ relates two Mn spins along the $c$-axis and has a character of 1. For a  $b$-axis Mn magnetic moment, this symmetry operator flips the spin, resulting in an antiferro arrangement along the $c$-axis (odd integer $l$). On the other hand, for a $c$-axis component, this same operator leaves the spin invariant (ferro along $c$ and peaks at even integer $l$). Therefore in a canted structure with Fourier components present at odd and even integer $l$ values, these must see the $b$ axis and $c$ -axis moment directions, respectively - as is observed.

Turning now to the low temperature structure, we again start from the the model of Kenzelmann {\it et al.,}\cite{Kenzelmann:2005p19}, who found a single Fourier component of $(0,\tau,1)$ with both $\Gamma_2$ and $\Gamma_3$ components with moment directions along the $c$-axis and $b$-axis respectively. We add to this the $(0,\tau,0)$ $c$-axis component, as in the high temperature phase, and the  $(0,1-\tau,0)$ $a$-axis Fourier component. The fact that this latter reflection is a satellite of the \miller{0}{1}{0} \emph{forbidden} Bragg reflection, implies that this component is arranged antiferromagnetically along the $b$-axis. Thus this $(0,1-\tau,0)$ $a$-axis component causes a canting of the Mn moments towards the $a$-axis that is antiferro ordered along the $[010]$ direction. From the ratio of the azimuthal maxima of the \miller{0}{2\tau}{0} and \miller{0}{1-2\tau}{0} intensities we estimate the ratio of the $a$-axis component to the $c$-axis component to be $\sim 0.66$. These $a$-axis and $c$-axis Fourier components then modify the cycloid of Kenzelmann {\it et al.} and the resulting structure is shown in Fig.~\ref{fig:cartoon}(b).

The addition of the $(0,1-\tau,0)$ component is consistent with a $\Gamma_2$ representation. The Mn spins along the $b$-axis are related by the $b$-glide ($m_{yz}$) 
and the two-fold screw axis $2_y$.
In the $\Gamma_2$ representation, the characters of these operators are -1 and 1 respectively. For an $a$-axis spin, then, $\Gamma_2$ will have an anti-parallel alignment to $b$-axis neighbors and therefore a wavevector of the \miller{0}{1-\tau}{0} type. For the \miller{0}{\tau}{0} $c$-axis component, the same analysis as was discussed above in the high-temperature phase also applies below 28~K and it is a $\Gamma_3$ component. Thus this new low temperature structure is also consistent with Kenzelmann {\it et al.} with both $\Gamma_2$ and $\Gamma_3$ components.

We note the \miller{0}{1-2\tau}{0} reflection is also present in the high-temperature phase. While the intensity is too weak to carry out azimuthal analysis of this reflection, symmetry arguments suggest that it remains an $a$-axis component, in which case it would be a $\Gamma_2$ component and both representations would be present above 28 K too.

In conclusion, we propose new magnetic structures above and below 28~\kelvin\ in TbMnO$_3$, on the basis of soft x-ray resonant scattering measurements. Specifically, our data require the addition of $a$-axis and $c$-axis components, with Fourier components $(0,1-\tau,0)$ and $(0,\tau,0)$ respectively. These new components are present in both the high temperature and low temperature phases and mean that the ferroelectrically relevant transition at 28 K is not, as previously reported, from a collinear structure to non-collinear one, but rather both phases are non-collinear and both are described by $\Gamma_2$ and $\Gamma_3$ irreducible representations. The essential difference then in the ferroelectric phase is the cycloidal component to the magnetic structure that forms in the $\Gamma_2$ component. We emphasize that the new components do not change the direction of the ferroelectric polarization as given by Eq.~\ref{eq:P}. Our magnetic structure reconciles all observed wavevectors into a single coherent structure, and therefore disagrees with a model of coexistence, as suggested by Kajimoto {\it et al.,}\cite{Kajimoto:2004p1616} and a model of domains as suggested by Mannix {\it et al.}\cite{Mannix:2007p1566}.

The authors would like to thank C. Detlefs, B. Detlefs,  A. Wills, W. Ku, E. S. Bo\v{z}in and S.J. Billinge for many stimulating discussions. The work at Brookhaven National Laboratory is supported by the Office of Science, US Department of Energy, under contract no. DE-AC02-98CH10886. Work at UCL was supported by the EPSRC and a Wolfson Royal Society Award and in Durham and Oxford by the EPSRC.

\bibliography{sbwbib}
\end{document}